\documentclass[12pt]{article}
\begin{document}

\begin{flushright}
UT-Komaba 00-10  \\
\end{flushright}


\begin{center} 
{\Large{\bf  Lattice QCD with the Overlap Fermions at Strong
Gauge Coupling (II)}}

\vskip 1.5cm

{\Large  Ikuo Ichinose\footnote{e-mail 
 address: ikuo@hep1.c.u-tokyo.ac.jp}and 
 Keiichi Nagao\footnote{e-mail
 address: nagao@hep1.c.u-tokyo.ac.jp}}  
\vskip 0.5cm
 
Institute of Physics, University of Tokyo, Komaba,
  Tokyo, 153-8902 Japan  
 
\end{center}

\vskip 3cm
\begin{center} 
\begin{bf}
Abstract
\end{bf}
\end{center}
In the previous paper we developed a strong-coupling expansion
for the lattice QCD with the overlap fermions and showed that
L\"uscher's ``extended" chiral symmetry is spontaneously broken
in some parameter region of the overlap fermions.
In this paper, we derive a low-energy effective action of hadrons and 
show that there exist quasi-Nambu-Goldstone bosons which are
identified as the pions.
The pion field is a {\em nonlocal} composite field of quark 
and anti-quark even at 
the strong-coupling limit
because of the nonlocality of the overlap fermion formalism and
L\"uscher's chiral symmetry.
The pions become massless in the limit of the vanishing bare-quark mass
as it is desired. 
We furthermore examine symmetries of the overlap fermions with
the hopping expansion and argue that there appear no other massless
modes other than the pions.

\newpage
\setcounter{footnote}{0}
\section{Introduction}

One of the long standing problems in the lattice gauge theory
is the formulation of lattice fermions.
Recently a very promising formulation named overlap fermion
was proposed by Narayanan and Neuberger \cite{NN,Ne} and it has 
been studied intensively by both analytic and numerical methods.
In the previous paper \cite{IN} (which we shall refer to paper I hereafter),
we slightly extended the overlap 
fermion by introducing a ``hopping" parameter $t$ and studied
the lattice QCD by using both the $t$-expansion and the strong-coupling
expansion.
There we calculated the effective potential of the chiral condensation and
showed that L\"uscher's extended chiral symmetry \cite{Lus}
is spontaneously broken at certain parameter region of the overlap fermion.
In this paper we shall derive an effective action of low-energy excitations
and show that there exist quasi-Nambu-Goldstone bosons which are identified
as the pions.
As we show, the pion field is a {\em nonlocal} composite field 
of quark and anti-quark
even at the strong-coupling limit.

This paper is organized as follows.
In Sect.2, we shall give a brief review of the overlap fermion
and its hopping expansion.
We introduce nonlocal composite operators of quarks which transform
``covariantly" under L\"uscher's chiral transformation.
In Sect.3, we derive the effective action of mesons by integrating out
the gauge fields and the quarks.
L\"uscher's symmetry is spontaneously broken at strong gauge coupling
as we showed in paper I.
From the effective action, we show that
there appear quasi-massless mesons which are identified as the pions.
In Sect.4, we discuss symmtries of the overlap fermion and their
relation to the species doubling.
We argue that there are no massless modes other than the pions.
Physical meaning of the hopping expansion is also discussed.
Section 5 is devoted for discussion and conclusion.


\setcounter{equation}{0}
\section{Hopping expansion} 

Action of the overlap fermion on the $d$-dimensional lattice
is given as follows,
\begin{equation}
S_F=a^d\sum_{n,m}\bar{\psi}(m)D(m,n)\psi(n),
\label{SF}
\end{equation}
where the covariant derivative $D(m,n)$ is defined as
\begin{eqnarray}
D&=&{1\over a}\Big(1+X{1 \over \sqrt{X^{\dagger}X}}\Big),  \nonumber  \\
X_{mn}&=&\gamma_{\mu}C_{\mu}(t;m,n)+B(t;m,n),  \nonumber   \\
C_{\mu}(t;m,n)&=&{t \over 2a}\Big[\delta_{m+\mu,n}U_{\mu}(m)-
\delta_{m,n+\mu}U^{\dagger}_{\mu}(n)\Big],  \nonumber  \\
B(t;m,n)&=&-{M_0\over a}+{r\over 2a}\sum_{\mu}\Big[2\delta_{n,m}
-t \delta_{m+\mu,n}U_{\mu}(m)-t \delta_{m,n+\mu}U^{\dagger}_{\mu}(n)\Big],
\label{covD}
\end{eqnarray} 
where $r$ and $M_0$ are 
dimensionless nonvanishing free parameters of the overlap lattice fermion 
formalism and $U_{\mu}(m)$ is gauge field on links.
Other notations are standard.
We have introduced a new parameter $t$.\footnote{As we explained in
paper I, the $t$-dependence of the operator $D(m,n)$ is absorbed
by a redefinition of $M_0$. However introducing the parameter
$t$ makes the calculations below quite transparent.}
The original overlap fermion corresponds to $t=1$.
For notational simplicity, we define 
\begin{equation}
A\equiv {1\over a}(dr-M_0), \;\; 
B\equiv {rt \over 2a},  \;\;
C\equiv {t\over 2a},
\label{ABC}
\end{equation}
and 
\begin{eqnarray}
\Gamma^-_\mu (m,n)&\equiv&\delta_{m+\mu,n}U_{\mu}(m)-
\delta_{m,n+\mu}U^{\dagger}_{\mu}(n),  \nonumber  \\
\Gamma^+_\mu(m,n)&\equiv&\delta_{m+\mu,n}U_{\mu}(m)+
\delta_{m,n+\mu}U^{\dagger}_{\mu}(n).
\label{-+}
\end{eqnarray}
In terms of the above quantities,
\begin{equation}
X_{mn}=A\delta_{mn}+C\sum\gamma_\mu \Gamma^-_\mu (m,n)-
B\sum\Gamma^+_\mu (m,n),
\label{X2}
\end{equation}
\begin{equation}
(X^\dagger)_{mn}=A\delta_{mn}-C\sum\gamma_\mu \Gamma^-_\mu (m,n)-
B\sum\Gamma^+_\mu (m,n).
\label{X3}
\end{equation}
From Eq.(\ref{ABC}), $B,C =O(t)$ and we consider $A=O(1)$ in the later
discussion.
Then it is rather straightforward to expand $D(m,n)$ in powers of
$t$,
\begin{eqnarray}
aD(m,n)&=&2\theta(A)\delta_{mn}
+{C \over |A|}\sum \gamma_\mu\Gamma^-_\mu(m,n)  \nonumber   \\
&& \;\;
   +{BC \over 2A|A|}\sum \gamma_\mu\Big(\Gamma^-_\mu(m,l)
   \Gamma^+_\nu(l,n)+\Gamma^+_\nu(m,l)\Gamma^-_\mu(l,n)\Big)  \nonumber   \\
&& \;\; +{C^2 \over 2A|A|}\sum\gamma_\mu\gamma_\nu \Gamma^-_\mu(m,l)  
   \Gamma^-_\nu(l,n)+O(t^3).  
\label{tD}
\end{eqnarray} 
Higher-order terms of $t$ are nonlocal and the $t$-expansion corresponds to 
a kind of the hopping expansion.
In the free field case or at the weak gauge coupling, the parameter region in
which fermion propagator has no species doublers is easily identified
in the $(M_0,r)$ plane.
However in the strong-coupling theory like QCD, the parameter region
of physical relevance should be determined by other requirements, because
the pole in the quark propagator is {\em not} a physical observable.
Therefore it is important to study the lattice QCD with the overlap
fermions in rather wide region of the parameter space. 

For the negative $A$ case, which we are interested in later discussion,
the leading-order term of $aD(m,n)$ in Eq.(\ref{tD}) is nothing but
the naive fermion Dirac operator.
As well-known, the naive lattice fermion has larger symmetries than
the continuum Dirac fermion.
As a result, there appear massless modes other than the pions if
the chiral symmetry is spontaneously broken.
To avoid this, the terms up to at least $O(t^2)$ in Eq.(\ref{tD}) must be
taken for practical calculations. 
This point and its relation to the species doubling will be discussed 
rather in detail after obtaining the low-energy effective action of 
hadrons.\footnote{It has been conjectured that the (weak-coupling)
doubling problems might be irrelevant to strong-coupling confining
models if the extra symmetries of naive fermions are (explicitly)
broken. See for example Ref.\cite{slac}.}

It is verified that the Ginsparg-Wilson (GW) relation \cite{GW}
\begin{equation}
D\gamma_5+\gamma_5D=aD\gamma_5D,
\label{GW}
\end{equation}
is satisfied by the $t$-expanded $D(m,n)$ in Eq.(\ref{tD}) at each order of
$t$.
Action of the fermion $S_F$ in Eq.(\ref{SF}) is invariant under 
the following extended chiral transformation discovered
by L\"uscher \cite{Lus},
\begin{equation}
\delta\psi(m)=\epsilon\gamma_5\Big(\delta_{nm}-aD(m,n)\Big)\psi(n),\;\;
\delta\bar{\psi}(m)=\epsilon\bar{\psi}(m)\gamma_5,
\label{extended}
\end{equation}
where $\epsilon$ is an infinitesimal transformation parameter.

Total action of the lattice QCD is given by
\begin{eqnarray}
S_{tot}&=&S_G+S_{F,M},  \nonumber   \\
S_G&=&-{1\over g^2}\sum_{pl}\mbox{Tr}(UUU^\dagger U^\dagger),  \nonumber  \\
S_{F,M}&=& S_F-M_B\sum\bar{\psi}(m)\psi(m),
\label{Stot}
\end{eqnarray}
where we have added the bare mass term of quarks.
We shall consider the strong-coupling limit in this paper though
a systematic strong-coupling expansion is possible.
We consider the $U(N)$ gauge group for large $N$.
It is easily verified that the following composite operators are
covariant under the transformation (\ref{extended})\cite{Nieder,Kiku},
\begin{equation}
\bar{q}(n)\equiv \bar{\psi}(n), \;\; q(n)\equiv \Big(1-{a\over 2}D(n,m)\Big)
\psi(n),
\label{qs}
\end{equation}
that is
\begin{equation}
\delta q(m)={\epsilon}\gamma_5 q(m), \;\; 
\delta \bar{q}(m)=\epsilon \bar{q}(m)\gamma_5.
\label{chiralq}
\end{equation}
Hereafter we often set the lattice spacing $a=1$.
We consider the case of negative $A$ which is expected to
have desired properties of QCD \cite{IN}.

\setcounter{equation}{0}
\section{Effective action of hadrons}

Partition function of the $U(N)$ QCD is given by the following functional
integral,
\begin{equation}
Z[J]=\int D\bar{\psi}D\psi DU \exp \Big\{-S_{tot}+\sum J(n)\hat{Q}(n)\Big\},
\label{partition}
\end{equation}
where $[DU]$ is the Haar measure and
\begin{eqnarray}
&& J(n)\hat{Q}(n)=J^\alpha_\beta(n)\hat{Q}_\alpha^\beta(n),  \nonumber  \\
&&  \hat{Q}_\alpha^\beta(n)=
{1\over N}\sum_aq_{a,\alpha}(n)
\bar{q}^{a,\beta}(n),
\label{Jm}
\end{eqnarray}
with color index $a$ and spinor-flavor indices $\alpha$ and $\beta$.
It should be remarked that the source $J$ is coupled to the nonlocal
operator $\hat{Q}$ instead of $\sum_a\psi_{a,\alpha}(n)
\bar{\psi}^{a,\beta}(n)$.
In Ref.\cite{IN-GN} we studied the gauged Gross-Neveu model
with the overlap fermions and showed that $\hat{Q}_\alpha^\beta(n)$
is the proper composite fields for the extended chiral symmetry, i.e.,
the order parameter is given by $\hat{Q}_\alpha^\beta(n)\delta_{\alpha\beta}$
and the Nambu-Goldstone bosons correspond to tr$_S(\hat{Q}(n)\gamma_5)$
where tr$_S$ is the trace over spinor indices.
In the rest of this section, we shall derive the effective 
action $S_{eff}({\cal Q})$ defined as
\begin{equation}
Z[J]=\int D{\cal Q} e^{-S_{eff}({\cal Q})+J{\cal Q}}
\label{Seff}
\end{equation}
where integral over color-singlet {\em elementary} ``meson" field 
${\cal Q}^\alpha_\beta$
is defined as in Ref.\cite{IN}.

In order to evaluate the partition function (\ref{partition}),
we make a change of variables as
$$
(\bar{\psi},\psi) \Rightarrow  (\bar{q},q).
$$
Then the measure of the functional integral is transformed as
\begin{equation}
[d\bar{\psi}d\psi] \Rightarrow [d\bar{q}dq]e^{N{\rm Tr}\ln (1-{a\over2}D)},
\label{psi-q}
\end{equation}
where Tr in (\ref{psi-q}) is the trace over the spinor-flavor 
as well as the real-space indices. 
The contribution from the Jacobian $\mbox{Tr}\ln (1-{a\over2}D)$
is easily evaluated by the $t$-expansion.
In terms of $\bar{q}$ and $q$, 
\begin{eqnarray}
S'_{F}(q)&=&S_{F}(\psi) \nonumber \\
&=&\bar{q}(m)\Bigg[{C \over |A|}\sum \gamma_\mu\Gamma^-_\mu(m,n) 
+{BC \over 2A|A|}\sum \gamma_\mu\Big(\Gamma^-_\mu(m,l)
   \Gamma^+_\nu(l,n)+\Gamma^+_\nu(m,l)\Gamma^-_\mu(l,n)\Big)  \nonumber   \\
&& +O(t^3)\Bigg]q(n).
\label{Sq}
\end{eqnarray}
The term of $O((C/A)^2)$ in (\ref{tD}) has disappeared. As explained in 
paper I, this term is exactly determined from the term of $O(C/A)$
in (\ref{tD}) through the GW relation.
We expect that a similar phenomenon occurs for higher-order terms
of the $t$-expansion
which are determined by lower-order terms through the GW relation. 
That is, the terms of higher-order of $t$ which are determined 
through the GW relation 
drop when the action is rewritten in terms of $q$ and $\bar{q}$.
The mass term is also given by
\begin{eqnarray}
M_B\bar{\psi}(m)\psi(m)&=&M_B\bar{q}(m)\Big[\delta_{mn}-
{C \over 2A}\sum \gamma_\mu\Gamma^-_\mu(m,n) \nonumber  \\
&&-{BC \over 4A^2}\sum \gamma_\mu\Big(\Gamma^-_\mu(m,l)
   \Gamma^+_\nu(l,n)+\Gamma^+_\nu(m,l)\Gamma^-_\mu(l,n)\Big) \nonumber \\
&&   +O(t^3)\Big]q(n).
\end{eqnarray}
Total action of $q$ is then given by
\begin{equation}
S'_{F,M}(q)=S_{F,M}(\psi).
\end{equation}

Integral over the gauge field can be performed by the one-link integral,
\begin{equation}
e^{W(\bar{D},D)}=\int dU_\mu \exp\Big[\mbox{Tr}(\bar{D}_\mu U_\mu
+U^\dagger_\mu D_\mu)\Big],
\label{onelink}
\end{equation}
where $D_\mu$ and $\bar{D}_\mu$ are bilinears of Grassmann numbers.
Explicit form of the one-link integral $W(\bar{D},D)$ for the
present system is obtained as in paper I.
Gauge fields in $\Gamma^\pm_\mu(m,n)$ are replaced with composite
operators of $q$ and $\bar{q}$ after the integral over $U_\mu(n)$.
(For details see paper I.)
After some calculation, 
\begin{equation}
\int DUe^{-S'_{F,M}(q)+N{\rm Tr}\ln (1-{1\over2}D)+\sum J\hat{Q}}
=e^{-NM_B\mbox{tr}(\hat{Q})-S_2(\hat{Q})+\sum J\hat{Q}},
\label{U-int}
\end{equation}
where tr in (\ref{U-int}) is the trace over spinor-flavor indices and 
$S_2(\hat{Q})$ is some complicated function of $\hat{Q}$ which is obtained
in powers of $t$.
Let us define the following operators;
\begin{eqnarray}
\epsilon&=&\epsilon_\mu^{\delta\sigma}(m)=\Big({C \over A}\Big)^2
\Big(1-{M_B\over 2}\Big)^2\Big(\hat{Q}(m)\gamma_\mu\hat{Q}(m+\mu)
\gamma_\mu\Big)^{\delta\sigma},  \nonumber  \\
\epsilon' &=&\epsilon_\mu^{'\delta\sigma}(m)=\Big({C \over A}\Big)^2
\Big(1-{M_B\over 2}\Big)^2\Big(\hat{Q}(m+\mu)\gamma_\mu\hat{Q}(m)
\gamma_\mu\Big)^{\delta\sigma}.
\label{epsilons}
\end{eqnarray}
Then in terms of $\epsilon$'s, $S_2(\hat{Q})$ is given as\footnote{As 
we shall see, ${\cal Q} \sim O(t^{-1})$.} 
\begin{eqnarray}
{1\over N} S_2(\hat{Q})&=&-\sum_{m,\mu}\mbox{tr}
\Big[g(\epsilon_\mu(m))\Big]\nonumber \\
&&+{BC^3\over 2A^4}\Big(1-{M_B \over 2}\Big)^3 \nonumber  \\
&&\times \sum_{m,\mu,\nu}
\Bigg\{\mbox{tr}\Big[{\cal Q}(m+\mu)\gamma_\mu g'(\epsilon_\mu(m))
{\cal Q}(m)\gamma_\mu{\cal Q}(m+\mu+\nu)\gamma_\nu g'(\epsilon_\nu
(m+\mu))\Big]  \nonumber  \\
&&-\mbox{tr}\Big[{\cal Q}(m+\mu+\nu)\gamma_\mu g'(\epsilon_\mu(m+\nu))
{\cal Q}(m+\nu)\gamma_\mu{\cal Q}(m+\mu)\gamma_\nu g'(\epsilon'_\nu
(m+\mu))\Big]  \nonumber  \\
&&+\mbox{tr}\Big[{\cal Q}(m)\gamma_\mu g'(\epsilon'_\mu(m))
{\cal Q}(m+\mu)\gamma_\mu{\cal Q}(m+\nu)\gamma_\nu g'(\epsilon_\nu
(m))\Big]   \nonumber  \\
&&-\mbox{tr}\Big[{\cal Q}(m+\nu)\gamma_\mu g'(\epsilon'_\mu(m+\nu))
{\cal Q}(m+\mu+\nu)\gamma_\mu{\cal Q}(m)\gamma_\nu g'(\epsilon'_\nu
(m))\Big]  \nonumber  \\
&&+\mbox{tr}\Big[{\cal Q}(m+\nu)\gamma_\nu g'(\epsilon_\nu(m))
{\cal Q}(m)\gamma_\mu{\cal Q}(m+\mu+\nu)\gamma_\mu g'(\epsilon_\mu
(m+\nu))\Big]   \nonumber  \\
&&+\mbox{tr}\Big[{\cal Q}(m+\mu+\nu)\gamma_\nu g'(\epsilon_\nu(m+\mu))
{\cal Q}(m+\mu)\gamma_\mu{\cal Q}(m+\nu)\gamma_\mu g'(\epsilon'_\mu
(m+\nu))\Big]  \nonumber   \\
&&-\mbox{tr}\Big[{\cal Q}(m)\gamma_\nu g'(\epsilon'_\nu(m))
{\cal Q}(m+\nu)\gamma_\mu{\cal Q}(m+\mu)\gamma_\mu g'(\epsilon_\mu
(m))\Big]  \nonumber  \\
&&-\mbox{tr}\Big[{\cal Q}(m+\mu)\gamma_\nu g'(\epsilon'_\nu(m+\mu))
{\cal Q}(m+\mu+\nu)\gamma_\mu{\cal Q}(m)\gamma_\mu g'(\epsilon'_\mu
(m))\Big] \Bigg\} \nonumber  \\
&&+{N_{sf} C^4\over 4A^4}\Big(1-{M_B \over 2}\Big)^2
\sum_{m,\mu}\mbox{tr}\Big[{\cal Q}(m+\mu)\gamma_\mu g'(\epsilon_\mu(m))
{\cal Q}(m)\gamma_\mu g'(\epsilon'_\mu
(m))\Big]  \nonumber   \\
&& +O(t^3), 
\label{S2}
\end{eqnarray}
where $N_{sf}$ is the dimension of the spinor-flavor index and 
\begin{equation}
g(x)=1-(1-4x)^{1\over 2}+\ln \Big[{1\over 2}(1+(1-4x)^{1\over 2})\Big].
\label{g(x)}
\end{equation}

Elementary meson fields ${\cal Q}$ and their functional integral 
are introduced as in the previous case \cite{IN,KS},
\begin{eqnarray}
\int d\bar{q}dq \exp\Big({1\over N}J^\beta_\alpha q^\alpha_a
\bar{q}^a_\beta\Big) &=& \Big(\mbox{det}J\Big)^N  \nonumber   \\
&=&\oint d{\cal Q}
\Big(\mbox{det}{\cal Q} \Big)^{-N}\cdot e^{J\cdot{\cal Q}},
\label{int-meson}
\end{eqnarray}
where the integral over ${\cal Q}$ is defined by the contour integral,
i.e., ${\cal Q}$ is polar-decomposed as ${\cal Q}=RV$ with positive-definite
Hermitian matrix $R$ and unitary matrix $V$, and 
$\oint d{\cal Q}\equiv \int dV$ with the Haar measure of U($N_{sf}$)\cite{KS}.
From (\ref{int-meson}), there appear additional terms like
$(N\mbox{Tr}\log {\cal Q})$ in the effective action.
Therefore the effective action is given by
\begin{equation}
S_{eff}({\cal Q})=N\sum_n\Big[\mbox{tr} \ln {\cal Q}(n)
+M_B \mbox{tr} {\cal Q}(n)\Big]
+S_2({\cal Q}).
\label{Seff2}
\end{equation}

Effective potential of the chiral condensate is obtained from
$S_{eff}$ in Eq.(\ref{Seff2}) by setting 
$$
Q^{\alpha\beta}(n)=v\delta_{\alpha\beta}.
$$
In paper I we obtained
\begin{equation}
v={|A| \over 2C}\sqrt{{2d-1 \over d^2}} +O(t^0)+O(M_B).
\label{VEV2}
\end{equation}
From the above result, we can expect that there appear quasi-Nambu-Goldstone
bosons, i.e., pions.
Actually we found that there are massless modes in the channel 
$\mbox{tr}_S\Big(\hat{Q}\gamma_5\Big)=\mbox{tr}_S\Big(q\bar{q}\gamma_5\Big)$
where $\mbox{tr}_S$ is the trace over spinor index.
It is straightforward to obtain the effective action of the pions
by inserting the following expression of ${\cal Q}$
into $S_{eff}$ in Eq.(\ref{Seff2}),
\begin{equation}
{\cal Q}(m)=ve^{i\gamma_5\phi_5(m)}.
\label{phi5}
\end{equation}
For example from (\ref{epsilons}) and (\ref{phi5}), one can easily obtain
\begin{equation}
\epsilon_\mu \propto e^{-i\gamma_5\nabla_\mu\phi_5}.
\end{equation}
After some calculation, we have
\begin{equation}
S_{eff}|_{{{\cal Q}(m)=ve^{i\gamma_5\phi_5(m)}}}=
2^{\frac{d}{2}}N \Big[ C_\pi
\sum_{m,\mu}\mbox{tr}_F\Big(\nabla_\mu\phi_5(m)\Big)^2
-\frac{M_B v}{2} \sum_m\mbox{tr}_F\Big(\phi_5(m)\Big)^2 \Big],
\label{S-pion}
\end{equation}
where $C_\pi$ is some positive constant and $\mbox{tr}_F$ is the trace
over flavor index .
Therefore the fields $\phi_5$ are quasi-Nambu-Goldstone pions as expected.

One may think that there appear additional massless excitations
in other modes of ${\cal Q}(m)$ as in the naive lattice fermion case\cite{KS}.
In the following section we shall argue that this is not the case.
There relationship between the species doubling and the symmetries 
of the lattice fermion becomes clear.

\setcounter{equation}{0}
\section{Symmetries and species doubling}

In paper I we showed that L\"uscher's chiral symmetry is 
spontaneously broken at strong coupling.
In the previous section, we derived the effective action
and showed that pions appear in the desired form.
As well-known, the naive fermion has much larger symmetry than the 
continuum fermion, and as a result there appear massless vector boson,
etc. besides pions.
In this section, we shall examine symmetries of the $t$-expanded
overlap fermions and argue that such extra ``Nambu-Goldstone bosons"
do not appear because of the higher-order terms of $t$.
In that discussion, we show that these extra (quasi-)massless bosons are
nothing but pion like bound state of {\em quark and its doublers}.
Finally we shall discuss physical meaning of the hopping expansion
especially in the confining phase.

The naive lattice fermion action has ``spectrum doubling symmetry", that is, 
the action is invariant under the following transformation,
\begin{eqnarray}
\psi(n) &\rightarrow& e^{-in\cdot\pi_H} M_H \psi(n), \nonumber\\
\bar\psi(n) &\rightarrow& \bar\psi(n) M_H^\dag  
e^{in\cdot\pi_H}, \label{doubling}   \\  
\mbox{where} &&  n\cdot\pi_H=\sum_{\mu=1}^d n_\mu\pi_{H,\mu},  \nonumber\\ 
H&=& \{\mu_1,\cdot\cdot\cdot, \mu_h\}, 
\quad (\mu_1 < \mu_2 < \cdot\cdot\cdot < \mu_h,\; 1\leq h\leq d), \nonumber\\
\mbox{explicitly} && 
\{\mu\}=\{1\}, \cdots, \{d\}, \;
\{\mu_1,\mu_2\}=\{1,2\}, \{2,3\}, \{1,3\}, \cdots,  \nonumber \\
&& \{\mu_1,\mu_2, \cdots, \mu_d\}=\{1,2,\cdots,d\}, \nonumber \\
M_H&=&M_{\mu_1} M_{\mu_2}\cdot\cdot\cdot M_{\mu_h}, \nonumber\\
M_\mu &=& i\gamma_5 \gamma_\mu, \nonumber\\
\pi_{H,\mu} &=& \left\{
\begin{array}{ll}
\pi &  \mbox{if} \quad \mu \in H  \\
0 &  \mbox{otherwise}.
\end{array}
\right.
\end{eqnarray}
For example, under the transformation $H=\{\mu\}$, 
right(left)-handed fermion at momentum $p=0$
transforms to left(right)-handed fermion at $p=(0,..., p_\mu=\pi,0,...)$.
Furthermore we can diagonalize the spinor indices of fermions 
by rewriting the naive-fermion action $S_{naive}$ in terms of
the following variables $\chi_n$ and $\bar\chi_n$,
\begin{eqnarray}
\psi(n) &=& A_n \chi_n, \nonumber \\
\bar\psi(n)  &=& \bar\chi_n A_n^\dag, \nonumber \\
A_n&=&\gamma_1^{n_1} \gamma_2^{n_2} \cdots \gamma_d^{n_d},\nonumber \\ 
A_n^\dag \gamma_\mu A_{n \pm \hat\mu} &=& (-1)^{n_1+
\cdots n_{\mu-1}}1\label{chi},
\end{eqnarray}
\begin{equation}
S_{naive}=\sum \bar{\chi}_m\eta_\mu (n) 
\Big[\delta_{m+\mu,n}U_{\mu}(m)-
\delta_{m,n+\mu}U^{\dagger}_{\mu}(n)\Big]\chi_n, \label{Snaive}
\end{equation}
where $\eta_\mu (n)={1 \over 2}(-)^{n_1+\cdots+ n_{\mu-1}}$
and we have set the lattice spacing $a=1$.

As spinor indices of $\chi_n$ and $\bar\chi_n$ are diagonal in (\ref{Snaive}), 
the naive fermion action has global $U(N_{sf})\otimes U(N_{sf})$ symmetry 
which is manifest in the 
$\chi$-representation \cite{KS}.
By the chiral condensation $\langle \bar{\chi}_n\chi_n\rangle \neq 0$, 
the $U(N_{sf})\otimes U(N_{sf})$ symmetry
is spontaneously broken down to $U(N_{sf})$.
On the other hand, one can easily see how the fields $\chi_n$ 
and $\bar{\chi}_n$ transform under (\ref{doubling}).
For example, for $H=\{\mu\}$, 
\begin{eqnarray}
\chi_n &\rightarrow& i\gamma_5 \gamma _{\mu} \chi_n, \nonumber\\
\bar\chi_n &\rightarrow& \bar\chi_n i\gamma_5 \gamma_{\mu},\label{chitrans} 
\end{eqnarray}
and similar expression for other doubling transformation $H$.
Then it is obvious that the doubling symmetry (\ref{doubling})
is a part of the $U(N_{sf})$ symmetry.
This indicates that the extra Nambu-Goldstone bosons are nothing but
pion like bound state of the original quark and its doublers (for more
details, see discussion below).
By practical calculation, we can verify that higher-order terms of $t$
in Eq.(\ref{tD}) explicitly breaks the 
$U(N_{sf})\otimes U(N_{sf})$ symmetry down to
the flavour symmetry $U(N_f)$ as desired.
This is due to the Wilson term $B(t;m,n)$ in Eq.(\ref{covD}).

Next let us study the doubling spectrum of pions, e.g.,
``pions" which are composed of quark $q(p\sim 0)$ and, e.g., anti-quark 
$\bar{q} (p_\mu\sim \pi,p_{\nu\neq \mu}\sim 0)$\footnote{As
quarks are confined, concept of ``quark with vanishing momentum" etc
does not have any strict meaning. However, 
we hope that what we are meaning in the present context
is quite clear for readers.}.
From the doubling transformation (\ref{doubling}), we should consider
the following composite fields,
\begin{equation}
\hat{Q}'(n)\equiv {1\over N}\sum q(n)\bar{q}(n) M_H^\dag e^{in\cdot\pi_H}.
\end{equation}
For example for $H=\{\mu\}$,
\begin{equation}
\hat{Q}'=\hat{Q}i\gamma_\mu\gamma_5(-)^{n_\mu}\label{Qprime}. \nonumber
\end{equation}
From the above consideration, the bound state of 
quark $q(p\sim 0)$ and anti-quark 
$\bar{q} (p_\mu\sim \pi,p_{\nu\neq \mu}\sim 0)$ is described by the 
``vector" field $\phi_\mu$ introduced as follows,
\begin{equation}
Q=v e^{i\gamma_\mu \phi_\mu (-1)^{n_\mu}} ,\label{phi_mu}
\end{equation}
where the flavour index is arbitrary.

It is rather straightforward to obtain the effective action of 
$\phi_\mu$ by substituting Eq.(\ref{phi_mu}) into the
effective action $S_{eff}(\cal{Q})$ and using the formulae like
\begin{eqnarray}
\epsilon_\mu (m)&=& (\frac{C}{A})^2 (1-\frac{M_B}{2})^2 v^2 
e^{-i(-1)^{m_\nu} \gamma_\nu \nabla_\mu \phi_\nu (m)},  \\
\epsilon_\mu (m)^\prime &=& (\frac{C}{A})^2 (1-\frac{M_B}{2})^2 v^2 
e^{i(-1)^{m_\nu} \gamma_\nu \nabla_\mu \phi_\nu (m)}.
\end{eqnarray}
It is verified that at the leading order of $t$, i.e., the naive fermion,
$\phi_\mu$ has a mass like (mass)$^2\propto M_B$ just as the pions
$\phi_5$.
At $O(t^2)$, no mass terms appear (accidentally).
We therefore examine the terms of $O(t^3)$ of the hopping expansion like 
\begin{equation}
\frac{B^2 C}{A^3} \sum \gamma_\mu \Gamma_\mu^- (m,l)\Gamma_\nu^+ (l,v)
\Gamma_\alpha^+ (v,n) + \cdot \cdot \cdot.
\label{t3}
\end{equation}
The terms in Eq.(\ref{t3}) generate the following terms 
in the effective action
$S_{eff}({\cal Q})$, 
\begin{eqnarray}
&&\sum_{m,\mu,\nu,\alpha}
\mbox{tr}\Big[{\cal Q}(m)\gamma_\mu {\cal Q}(m+\mu+\nu+\alpha)
\gamma_\alpha g'(\epsilon_\alpha(m+\mu+\nu)){\cal Q}(m+\mu+\nu) \nonumber \\
&& \;\times \gamma_\nu g'(\epsilon_\nu(m+\mu)){\cal Q}(m+\mu)\gamma_\mu 
g'(\epsilon_\mu(m))\Big]+\cdots.
\label{higher}
\end{eqnarray}
By substituting the expression (\ref{phi_mu}) into Eq.(\ref{higher}),
one can verify that the mass term of $\phi_\mu$, i.e., the term
like $\phi^2_\mu$, actually appears at $O(t^3)$.

Similar consideration is straightforward for other bound state
of quark and its doublers.
These excitations are described by the following modes,
\begin{equation}
Q=v e^{i\gamma_\mu\gamma_\nu\gamma_5\phi_{\mu\nu}(-)^{n_\mu+n_\nu}},
\end{equation}
etc.
We expect that they all acquire a finite
mass at higher order of $t$ as the symmetry of the action 
indicates.
If the above expectation is correct, the result implies that
the doublers at weak coupling do not produce any ill effect in the 
spectrum of light mesons
at strong coupling {\em except} the $U(1)$ problem, though the result
is obtained by using the $t$-expansion in the above study.
Recently, Golterman and Shamir gave careful study on QCD with
the overlap fermions with small hopping parameter and they obtained the result
which supports the above argument \cite{GS}.
Relationship between the $U(1)$ problem and the $t$-expansion and/or
the strong-coupling expansion
will be briefly discussed in Sect.5.  

So far we have considered bound state of pairs like 
quark $q(p\sim 0)$ and anti-quark 
$\bar{q} (p_\mu\sim \pi)$ and found that they are massive.
In the rest of this section, we shall reexamine the pions 
$\phi_5$ more precisely.
Pions $\phi_5$ consist of the bound state of quark $q(p\sim 0)$
and anti-quark $\bar{q}(p \sim 0)$.
However besides that component, there are contributions from
other pairs of quark and anti-quark like $q(p_\mu\sim \pi)$ and
$\bar{q}(p_\mu \sim \pi)$, i.e., bound state of doublers
with the same momentum, i.e.,
\begin{equation}
\phi_5 \sim \bar{q}(p\sim 0)\gamma_5 q(p\sim 0)+\lambda\cdot
\bar{q}(p_\mu\sim \pi)\gamma_5 q(p_\mu\sim \pi)+\cdots.
\label{mixing}
\end{equation}
It is interesting to estimate the weight of these bound pairs $\lambda$.

To this end, let us first consider the weak-coupling region
or small lattice spacing and $U_\mu(m) \sim 1$.
From $S'_F(q)$ in (\ref{Sq}), the free propagator of $q$ is 
obtained from the following expression of the $t$-expanded action,
\begin{equation}
S'_F(q) \sim  \bar{q}(p)\Big[{C\over |A|}\sum \gamma_\mu 
i\sin p_\mu 
+{BC\over A|A|}\sum \gamma_\mu i\sin p_\mu \cos p_\nu\Big]
q(p)+\cdots,
\label{Sqp}
\end{equation}
where we have ignored coefficients which are irrelevant for the discussion.
From Eq.(\ref{Sqp}), for quark with momentum $p\sim 0$,
\begin{equation}
S'_F(q) \sim  \bar{q}(p)\Big({C\over |A|}+{BCd\over A|A|}\Big)
\sum\Big( i\gamma_\mu p_\mu\Big) q(p),\label{p-0}
\end{equation}
whereas for doubler at $p_\nu \sim \pi$ the second term in (\ref{Sqp})
changes sign because of the factor $\cos p_\nu$, and therefore 
\begin{equation}
S'_F(q) \sim  \bar{q}(p)\Big({C\over |A|}+{BC(d-2)\over A|A|}\Big)
\sum\Big( i\gamma_\mu p_\mu\Big) q(p), \label{p-pi}
\end{equation}
with the suitable redefinition of the momentum 
$p_\nu\rightarrow \pi-p_\nu$.
This means that at the leading order of the hopping expansion
the doublers have the same weight or residue with the original 
fermion at $p\sim 0$.
However at $O(t^2)$, residue of doublers is {\em reduced} though the fermion
propagator has a pole at the doubler momentum\footnote{Here
we assume negative $A$ as in the previous discussions.}.
That is, in the hopping expansion (for the free field case), 
the doublers appear as a genuine particle but their probability 
is decreased by the higher-order terms of $t$.
On the other hand, exact form of the free fermion
 propagator is obtained readily.
Physical parameter region of the overlap fermions is identified by
requiring that the propagator has a pole only at $p=0$ and no poles
at $p=\pi$.
In this sense, the $t$-expansion is not a good approximation
in the perturbative deconfinement phase.
However in the parameter region in which the $t$-expansion 
converges, we can see that doublers are suppressed in the overlap
formalism.

The above discussion can be easily extended for smooth gauge
field configurations or small gauge coupling $g$.
In this case, there appears an additional term proportional to the
gauge potential $igA_\mu =\ln U_\mu$ which is common to fermion
with small momentum and its doublers.
That is, both in Eqs.(\ref{p-0}) and (\ref{p-pi}),
\begin{equation}
\bar{q}(p)\Big(\gamma_\mu p_\mu\Big)q(p)
\rightarrow \bar{q}(k)\gamma_\mu\Big(\delta(k-p)p_\mu-gA_\mu(k-p)\Big)q(p).
\end{equation}
Then weights of doublers are suppressed by the term of $O(t^2)$
for smooth $A_\mu$ as in the free case.

In the confining phase on the other hand, poles in the quark
propagator are not physical observables.
We expect that the $t$-expansion gives at least qualitatively
correct picture for that phase.
The above consideration for the weak-coupling case suggests
that the weight $\lambda$ in Eq.(\ref{mixing}) is a decreasing function
of the hopping parameter $t$ for the probability of doublers
decreases as $t$ increases.
As $t$ increases further, convergence of the expansion breaks down.
We think that this nonconvergence of the $t$-expansion is closely
related with the $U(1)$ problem.
This point will be discussed in the following section.
At the strong-coupling limit, the convergence
of the $t$-expansion is improved and also the discrimination between
``quark with small momentum" and its doublers
becomes obscure because of the large fluctuation of the gauge field
$U_\mu(n)$. 
In order to clarify the phase structure of the QCD with overlap fermions,
numerical studies are required.
We hope that the studies by the $t$-expansion will be useful for
further studies on that problem.


\setcounter{equation}{0}
\section{Discussion}

In paper I, we developed the hopping expansion for the overlap fermions
and showed that L\"uscher's chiral symmetry
is spontaneously broken at strong gauge coupling.
In this paper, we derived the effective action of mesons
and showed that pions appear in the desired form.
It was also argued that other extra ``Nambu-Goldstone" bosons do
not appear because of the explicit breaking of the 
$U(N_{sf})\otimes U(N_{sf})$
symmetry $\rightarrow U(N_f)$ which exists in the naive lattice fermion.

Important problem is whether the $t$-expansion covers physically
relevant parameter region of the lattice QCD.
It is not straightfoward to answer this question because poles in the
quark propagator is not physical observable in the confinement phase.
However effect of the chiral anomaly must be appear in the 
physical parameter region.
In QCD, the $U(1)$ problem is a suitable object to identify the correct
parameter region.
We expect that a mass term
of the flavour singlet meson will appear from the Jacobian in 
(\ref{psi-q}).
This is a solution to the $U(1)$ problem.
However if this Jacobian ${\rm Tr}\ln (1-{1\over 2}D)$
and the action itself are both expandable in
powers of $t$, in other words, local in the real space in a broad sense, 
the total
action $S'_{F,M}(q)-N{\rm Tr}\ln (1-{1\over2}D)$ with $M_B=0$ is obviously
invariant under the transformation (\ref{chiralq}).
Therefore flavour singlet Nambu-Goldstone boson also appears
as a result of the spontaneous breaking of (\ref{chiralq}).
Recently careful study on the overlap-Dirac fermions with a small
hopping parameter is given \cite{GS}.
There the same result with above is obtained.

We expect that convergence of the $t$-expansion breaks down first
in the Jacobian because it contains (infinitely)
higher-power terms of the operator $D$.
Then a would-be Nambu-Goldstone boson becomes massive because
of the nonlocality\footnote{Under the chiral
transformation (\ref{chiralq}), the operator $D$ is invariant.
However after integration over the gauge field, the Jacobian
is expressed in terms of $q$. Then it is possible that symmetry-breaking 
terms appear there. In other words, the transformation from $\psi$
to $q$ becomes singular above a certain critical value of $t_C$.}.
Problem is whether this nonconvergence of the Jacobian factor
means a phase transition from small to intermediate (physical) value 
of $t$.
If this is a phase transition, 
results of the present study may be useless for investigation on the physical 
properties of the intermediate-$t$ phase.
Of course it depends on nature of the phase transition.
This problem is under study and we hope that result will be reported 
in a future publication \cite{WP}.

Finally let us notice recent studies on the domain wall fermions 
at strong coupling.
Very recently lattice $U(1)$ gauge model with the domain-wall fermions was
studied in the strong-coupling limit by the Hamiltonian 
formalism \cite{DW}.
There an effective Hamiltonian for low-lying color-singlet
degrees of freedom is obtained by treating the terms proportional
to the gauge fields $U_\mu(m)$ as perturbations.
This idea is very close to ours in paper I and the present paper
though we employ the Lagrangian formalism.
It is very interesting and also useful to investigate the relation 
between the results obtained in Ref.\cite{DW} and ours in detail.
This may give an important insights to numerical studies on QCD
with the domain-wall and /or overlap fermions.


\vskip 2cm
{\Large{\bf Acknowledgements}}

KN, one of the authors, would like to thank Prof.T.Yoneya for useful 
discussions.
We also thank the referee of this paper for pointing out the references
in Ref.\cite{slac}.


\end{document}